\documentstyle[nato,epsf]{crckapb}

\newcommand{\half}{\mbox{\small $\frac{1}{2}$}}          % 1/2
\newcommand{\third}{\mbox{\small $\frac{1}{3}$}}         % 1/3
\def\lsim{\mathrel{\rlap{\lower4pt\hbox{\hskip1pt$\sim$}}
    \raise1pt\hbox{$<$}}}                % less than or approx. symbol
\def\gsim{\mathrel{\rlap{\lower4pt\hbox{\hskip1pt$\sim$}}
    \raise1pt\hbox{$>$}}}                % greater than or approx. symbol

\begin{opening}
\title{QUENCHED QCD NEAR THE CHIRAL LIMIT$^{1}$}

\author{M.~G\"ockeler}
\author{P.~E.~L. Rakow}
\institute{Institut f\"ur Theoretische Physik,
           Universit\"at Regensburg, \\ D-93040 Regensburg, Germany}

\author{R.~Horsley}
\institute{Institut f\"ur Physik, Humboldt-Universit\"at zu Berlin,
           \\ D-10115 Berlin, Germany} 

\author{D.~Petters}
\author{D.~Pleiter}
\institute{Institut f\"ur Theoretische Physik,
           Freie Universit\"at Berlin,
           \\ D-14195 Berlin, Germany}

\author{G.~Schierholz}
\institute{Deutsches Elektronen-Synchrotron DESY \& NIC,
           \\ D-15735 Zeuthen, Germany}

\end{opening}

\begin{document}

\vspace{-12.50cm}                                           % for preprint
\noindent                                                   % for preprint
{\normalsize DESY 00--024}    \\                            % for preprint
{\normalsize HUB--EP--00/01}  \\                            % for preprint
{\normalsize TPR-00-03}       \\                            % for preprint
{\normalsize February 2000}   \\                            % for preprint
%\\ \\ \\                                                    % for talk
%\vspace{11.25cm}                                            % for 2 preprint
\vspace{10.625cm}                                            % for 2 preprint

% The \begin{document} command comes after the \end{opening}
% command.

%----------------------------------------------------------------------------

\begin{abstract}
A numerical study of quenched QCD for light quarks is presented using
$O(a)$ improved fermions. Particular attention is paid to
the possible existence and determination of quenched chiral logarithms.
A `safe' region to use for chiral extrapolations appears to be at
and above the strange quark mass%
\footnotetext[1]{Talk given by R. Horsley at the workshop 
``Lattice Fermions and the Structure of the Vacuum'', October 1999,
Dubna, Russia.}.                                            % for preprint
%\footnotetext[1]{Talk given by R. Horsley.}.                % for talk
\end{abstract}

%----------------------------------------------------------------------------

\section{Introduction}

The goal of lattice QCD is the
computation of physical quantities such as
hadron masses and matrix elements using numerical Monte Carlo methods.
This has proved to be an ambitious programme because
after discretisation of the path integral and generation
of a sufficiently large number of independent configurations
several limits must be considered:
\begin{enumerate}
   \item The box size. This is currently at $\sim 1.5 - 3 \, \mbox{fm}$,
         and should be compared with the nucleon {\it rms} radius
         of $\sim 0.8 \, \mbox{fm}$.
   \item The chiral limit $m_q \to 0$. The $u/d$ and $s$ quarks
         are light quarks.
   \item Continuum limit $a^k \to 0$
         ($k=2$ if we choose $O(a)$ Symanzik improved fermions,
         staggered fermions or Ginsparg-Wilson fermions; for Wilson
         fermions we expect the discretisation effects to have $k=1$).
\end{enumerate}
If all these limits can be successfully taken then presumably QCD
will reproduce nature. Although first attempts in this direction
are being made, \cite{alikhan99a}, it will require much faster computers
to achieve this goal. To reduce the computational effort
often the {\it quenched approximation}
is employed when the fermion determinant is simply set to a constant.
However then new problems arise (or are exacerbated):
\begin{enumerate}
   \item Spurious quenched chiral logarithms appear as $m_q \to 0$.
         \label{q_chiral_l}
   \item The appearance of {\it exceptional} configurations.
         \label{q_exceptional_l}
   \item Consistency of the final results when comparing with their
         experimental or phenomenological values.
         \label{q_cont_l}
\end{enumerate}
In this talk we shall consider points \ref{q_chiral_l} and
\ref{q_exceptional_l} numerically using $O(a)$ improved fermions.

%----------------------------------------------------------------------------

\section{Chiral perturbation theory and quenched chiral logarithms}

This was developed by Bernard and Golterman, \cite{bernard92a}
and Sharpe, \cite{sharpe92a}. What do we expect?
The quenched pseudoscalar effective chiral Lagrangian gives
\begin{eqnarray}
   (am_{ps})^2 
       &\propto& (a\widetilde{m}_q)^{1\over 1+\delta},
                                                     \nonumber  \\
   a^2 g_P
       &\equiv& \langle 0|\widehat{\cal P}| ps \rangle
                      \propto [(am_{ps})^2]^{-\delta},
                                                     \nonumber  \\
   am_{ps} af_{ps}
       &\equiv& \langle 0|\widehat{\cal A}_4 |ps\rangle 
                      \propto am_{ps}.
\label{decay_constant}
\end{eqnarray}
As the $\eta^\prime$ remains light and has a single and
double pole in its propagator, this latter term at $p^2 = 0$,
$(m_0^2/m_{ps}^2) (1/m_{ps}^2)$ acts like an extra vertex giving
a singular correction to the usually harmless loop term
$m_{ps}^2 \ln (m_{ps}/\Lambda^2)$ of $m_0^2 \ln (m_{ps}/\Lambda)^2$,
so that the logarithmic term becomes singular, \cite{gupta94a}.
This can then be summed to give eq.~(\ref{decay_constant}).
Normally PCAC, $\partial \cdot {\cal A} = 2 \widetilde{m}_q {\cal P}$,
would give us $(am_{ps})^2 \propto a\widetilde{m}_q$.
Thus the non-zero $\delta$
leads to singular behaviour for $m_{ps}$ and $g_P$ in the chiral limit
for quenched QCD. Simple estimates lead to an expectation for
$\delta \sim 0.1 - 0.2$. $\Lambda$ is a cut-off on the $\eta^\prime$ loop
so $\Lambda \lsim 900 \, \mbox{MeV}$ and above this scale any
singularities are surely damped out.
For consistency, we would expect little
difference between the PCAC quark mass, $a\widetilde{m}_q$, and
the `standard' quark mass, $am_q \equiv \half (1/\kappa - 1/\kappa_c)$.

Similarly by considering the vector/baryon
effective chiral Lagrangians, it can be shown that
\cite{booth96a,labrenz96a}
\begin{equation}
   am_{V,N} = c_0 + c_1 am_{ps} + c_2 (am_{ps})^2 + O((am_{ps})^3), 
\label{quenched_theory_fit}
\end{equation}
where there is an extra linear term present, as the $\eta^\prime$
gives to the usual term $(am_{ps})^3$ an additional term $m_0^2 am_{ps}$.

%----------------------------------------------------------------------------

\section{Numerical results}

\begin{figure}[htbp]
   \hspace*{0.75in}
   \epsfxsize=9.00cm
      \epsfbox{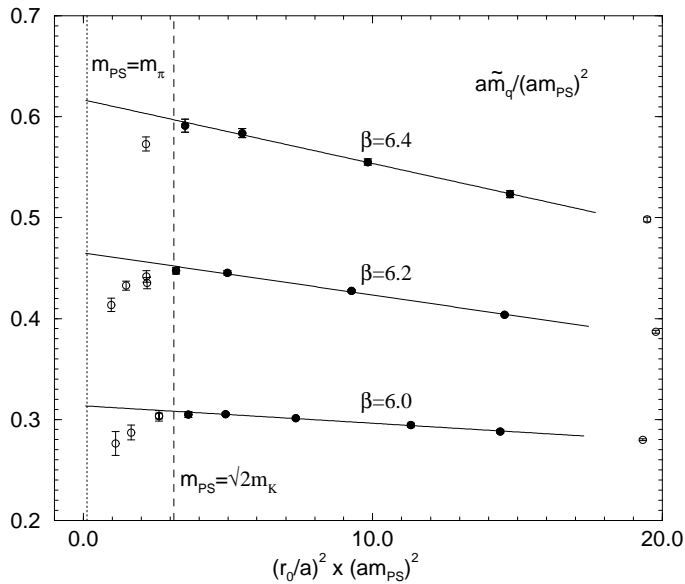}
   \caption{\it $a\widetilde{m}_q/(am_{ps})^2$ against $(am_{ps})^2$
            for $O(a)$ improved fermions.
            Filled circles denote points used in the linear fits.
            The dashed line is the mass of an (unphysical)
            $\overline{s}\gamma_5 s$ pseudoscalar meson, (using $m_K$),
            while the dotted line represents $m_\pi$.}
   \label{fig_mqwiompi2_mpi2_magic_lat99}
\end{figure}

How do these theoretical considerations fare with the numerical data?
Numerically, it advantageous to use the PCAC quark mass, $\widetilde{m}_q$
because this quark mass can be found very accurately
and does not depend on the first-to-be-determined parameter $\kappa_c$.
It is convenient to consider the ratio $a\widetilde{m}_q / (am_{ps})^2$.
This is shown in Fig.~\ref{fig_mqwiompi2_mpi2_magic_lat99},
\cite{gockeler99a}, for degenerate quark masses
at $\beta$ values of $6.0$, $6.2$ and $6.4$.
To give an idea of scales, we note that using the $r_0$ `force scale'
then $m_\pi$ lies almost at the chiral limit (within our
numerically accuracy there is no difference between these points)
and a hypothetical pseudoscalar meson, composed of the
strange quark and its antiquark,
lies at about $(r_0 m_{ps})^2 \sim 3.13$
(when using $m_K$). For the charm quark (using $m_D$)
we find $(r_0 m_{ps})^2 \sim 44.9$, way off the plot scale.

\begin{figure}[htbp]
   \hspace*{0.25in}
   \epsfxsize=11.00cm
   \epsfbox{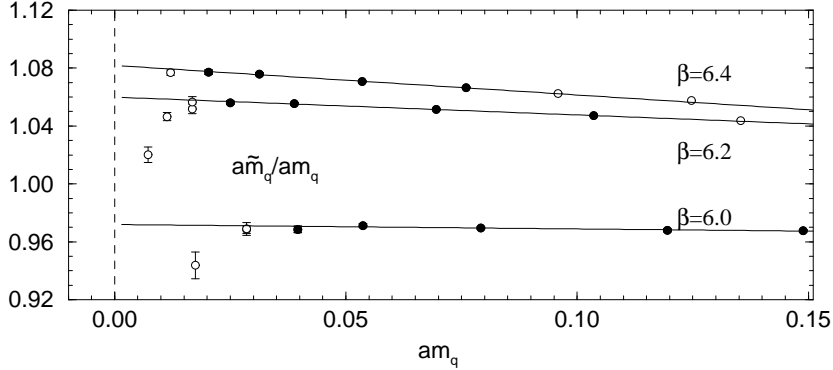}
   \caption{\it $a\widetilde{m}_q/am_q$ against $am_q$
            for $O(a)$ improved fermions.
            Filled circles denote points used in the fits.}
   \label{fig_ratio_magic_dubna99}
\end{figure}

It is to be seen from the picture that from
the strange quark mass to heavier quark masses
we have linear behaviour.
(Indeed the linearity seems to hold until
rather heavy quark masses, say $m_q \lsim \third m_c$.)
Below the strange quark mass,
there seems to be a (sharp?) break in this behaviour.
Possibly we can attribute this to the onset
of quenched chiral logarithms.
$m_{ps} = \sqrt{2}m_{K}$ corresponds here
to about $700 \, \mbox{MeV} \sim \Lambda$
so one might hope that any quenching effects are suppressed
above this value. We now check the behaviour
between $a\widetilde{m}_q$ and $am_q$.
In Fig.~\ref{fig_ratio_magic_dubna99}
we plot $a\widetilde{m}_q/am_q$ against $a m_q$.
While there seems to be a reasonably linear relation between the
two (lattice) definitions of the quark mass above $am_s$, below
we again see deviations. This fit is somewhat sensitive to
the value of $\kappa_c$ used; although the general picture
shown in Fig.~\ref{fig_ratio_magic_dubna99} never
seems to change significantly. (Indeed using all the light quark
data in the fit still produces a similar result.) 
This perhaps obscures the interpretation of
Fig.~\ref{fig_mqwiompi2_mpi2_magic_lat99} as being due to 
quenched chiral logarithms. Nevetheless, due to problems
in determining $am_q$, we prefer to use the results
with $a\widetilde{m}_q$, \cite{aoki99a}.

\begin{figure}[htbp]
   \hspace*{0.75in}
   \epsfxsize=9.00cm
   \epsfbox{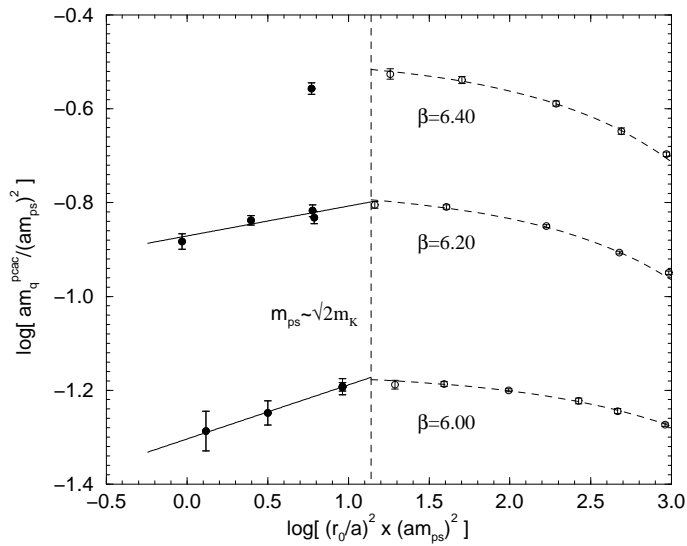}
   \caption{\it $\ln [(a\widetilde{m}_q / (am_\pi)^2]$ against
                $\ln [((r_0/a)^2 \times (am_{ps})^2]$.
                The left line is a linear fit to the quarks
                with mass below the strange quark mass, while
                the dotted line is the previous fit
                from Fig.~\ref{fig_mqwiompi2_mpi2_magic_lat99}.}
   \label{fig_log_mqwiompi2_mpi2_dubna99}
%   \vspace*{-0.25cm}
\end{figure}

To try to expose the small quark region, and to determine
$\delta$ (if the deviations are due to quenched chiral logarithms)
then it is convenient to plot the logarithm of
Fig.~\ref{fig_mqwiompi2_mpi2_magic_lat99}. This is done
in Fig.~\ref{fig_log_mqwiompi2_mpi2_dubna99},
where we expect the slope to be $\delta$ (for the
fitted quarks masses below the strange quark mass).
We find for $\beta = 6.0$, $\delta \sim 0.12(4)$, and
for $\beta = 6.2$, $\delta \sim 0.06(2)$ (for $\beta = 6.4$
there is not enough data). For $\beta = 6.0$, at least, 
there is reasonable agreement with the theoretical prejudice;
for $\beta =6.2$ the value seems small.

Despite the above results
it should be noted, as discussed above,
that it is notoriously difficult to
numerically detect quenched chiral logarithms.
Indeed the above effects may simply be due to finite-size effects
or `exceptional configuration' problems. We have only been
able to check very few quark mass points for finite
size effects. The impression is that they are small;
this is backed up by \cite{aoki99a}, who
work on a larger lattice.

\begin{figure}[htbp]
   \hspace*{0.25in}
   \epsfxsize=10.00cm
   \epsfbox{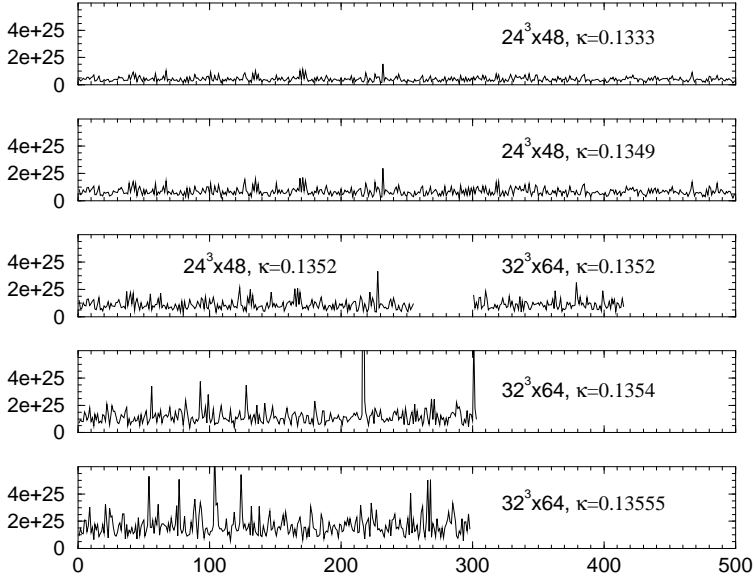}
   \caption{\it The pion norm, eq.~(\ref{pion_norm_def}),
                against configuration number for $\beta = 6.2$.}
   \label{fig_pion_norm_b6p20_dubna99}
\end{figure}

Exceptional configurations are seen as either the non-convergence
of the ferm\-ion matrix inversion or the correlator seems
to have a `fake source' at some $t$ value.
The problem is more severe for $O(a)$ improved fermions
than for Wilson fermions and increases
as $\beta \downarrow$ and/or $c_{sw} \uparrow$ and/or $m_q \downarrow$.
(This is the main obstacle for the $O(a)$ improvement programme
going below about $\beta \sim 6.0$.)
The reason for this problem is due to the presence of
small real eigenvalues in the fermion matrix.
In an experiment, \cite{gockeler98a},
we have chirally rotated the lattice quark mass, $am_q$,
away from the real axis; the same configuration then gave
a well behaved pion propagator. (We might then expect some mixing
in the correlation function of the particle with its parity
partner, however for the pion in the quark model this partner does
not exist.)
So perhaps simply throwing away the configuration does not
affect the spectrum (?). It is desirable
to have a crude indicator of whether we have an
exceptional configuration. In \cite{hoferichter95a}
the simple proposal was made to look at the pion norm,
\begin{equation}
   \Pi(\{U\}) = \sum_{\vec{x},t}
                  | \gamma_5 G(\vec{x},t ; \vec{0}, 0; \{U\}) \gamma_5 |^2.
\label{pion_norm_def}
\end{equation}
(In our application, the source $(\vec{0}, 0)$ was also Jacobi smeared.)
In Fig.~\ref{fig_pion_norm_b6p20_dubna99},
we show a sequence of pion norms for $\beta = 6.2$.
To decide on a criterion for an exceptional configuration
(ie a spike in the pion norm) is not so easy.
Some are obvious, for example from the pictures
we have at $\kappa = 0.1354$, $n_{conf}=217$ a problem.
(The corresponding pion propagator is shown in \cite{gockeler98a}.)
Closer to the critical point than here
more spikes have been seen in Wilson data, \cite{hoferichter99a}.
To be safe, we have actually chosen a more conservative local criterion
where if at any $t$ value the (pion) correlation function fluctuates
more than $5$ standard deviations from the local average we reject the
configuration. This leads at $\beta =6.2$, $32^3 \times 64$
for $\kappa = 0.1352$, $0.1354$, $0.13555$
to rejection rates of about $2$, $4$ and $6\%$ respectively.
(For the latter two $\kappa$ values about $15\%$ and
$33\%$ of these rates were actually due to non-convergence
of the inverter.)
So in conclusion: while we feel that for any
lighter quark mass than those considered here exceptional configurations
become a real problem, at our masses
while they are a nuisance, they do not distort
the numerical result.

\begin{figure}[htb]
   \hspace*{0.75in}
   \epsfxsize=9.00cm
   \epsfbox{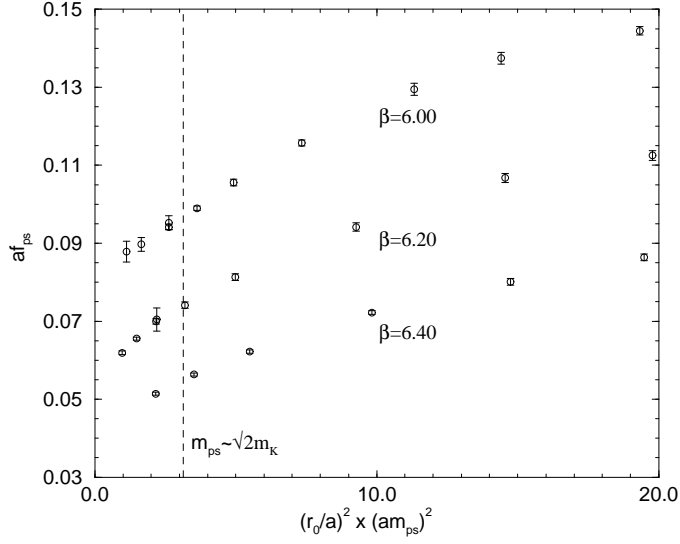}
   \caption{\it Unrenormalised (improved) $af_{ps}$
                versus the pseudoscalar mass.}
   \label{fig_f_pi_mpi2lin_dubna99}
\end{figure}

\begin{figure}[htb]
   \hspace*{0.75in}
   \epsfxsize=9.00cm
   \epsfbox{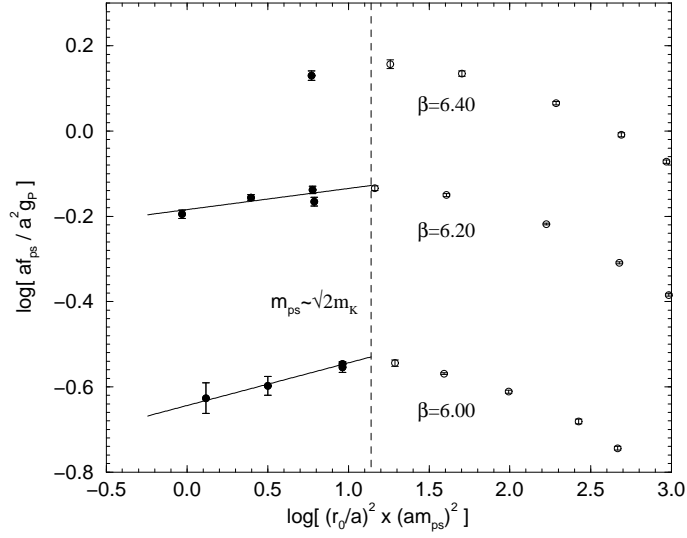}
   \caption{\it Unrenormalised (improved) $\ln [a f_{ps} / a^2 g_P ]$
                against $\ln [((r_0/a)^2 \times (am_{ps})^2]$.
                The left line is a linear fit to the results
                for quarks with mass below the strange quark mass.
                The same notation as for
                Fig.~\ref{fig_log_mqwiompi2_mpi2_dubna99}.}
   \label{fig_fg5ofpi_mpi2log_dubna99}
\end{figure}

We now turn to a consideration of the decay constant $f_{ps}$.
From eq.~(\ref{decay_constant}) we expect that $af_{ps}$ has
no quenched chiral singularity in it, while $a^2g_P$ diverges
in the chiral limit. In Fig.~\ref{fig_f_pi_mpi2lin_dubna99}
we plot the unrenormalised $af_{ps}$.
The results seem smooth over the whole quark mass range, with no singular
behaviour. Looking at the ratio $a f_{ps} / a^2 g_P$ we see the
same behaviour as in Fig.~\ref{fig_mqwiompi2_mpi2_magic_lat99},
ie for smaller quark masses than the strange quark mass a bend is
seen in the data.
As $af_{ps} / a^2 g_P \propto [ (am_{ps})^2 ]^\delta$
then taking the logarithm gives a direct estimate
of $\delta$. In Fig.~\ref{fig_fg5ofpi_mpi2log_dubna99}
we show this, with fit values
$\delta \sim 0.10(3)$ ($\beta = 6.0$) and
$\delta \sim 0.05(2)$ ($\beta = 6.2$) consistent with the
previous results.

\begin{figure}[htb]
   \hspace*{0.75in}
   \epsfxsize=8.50cm
   \epsfbox{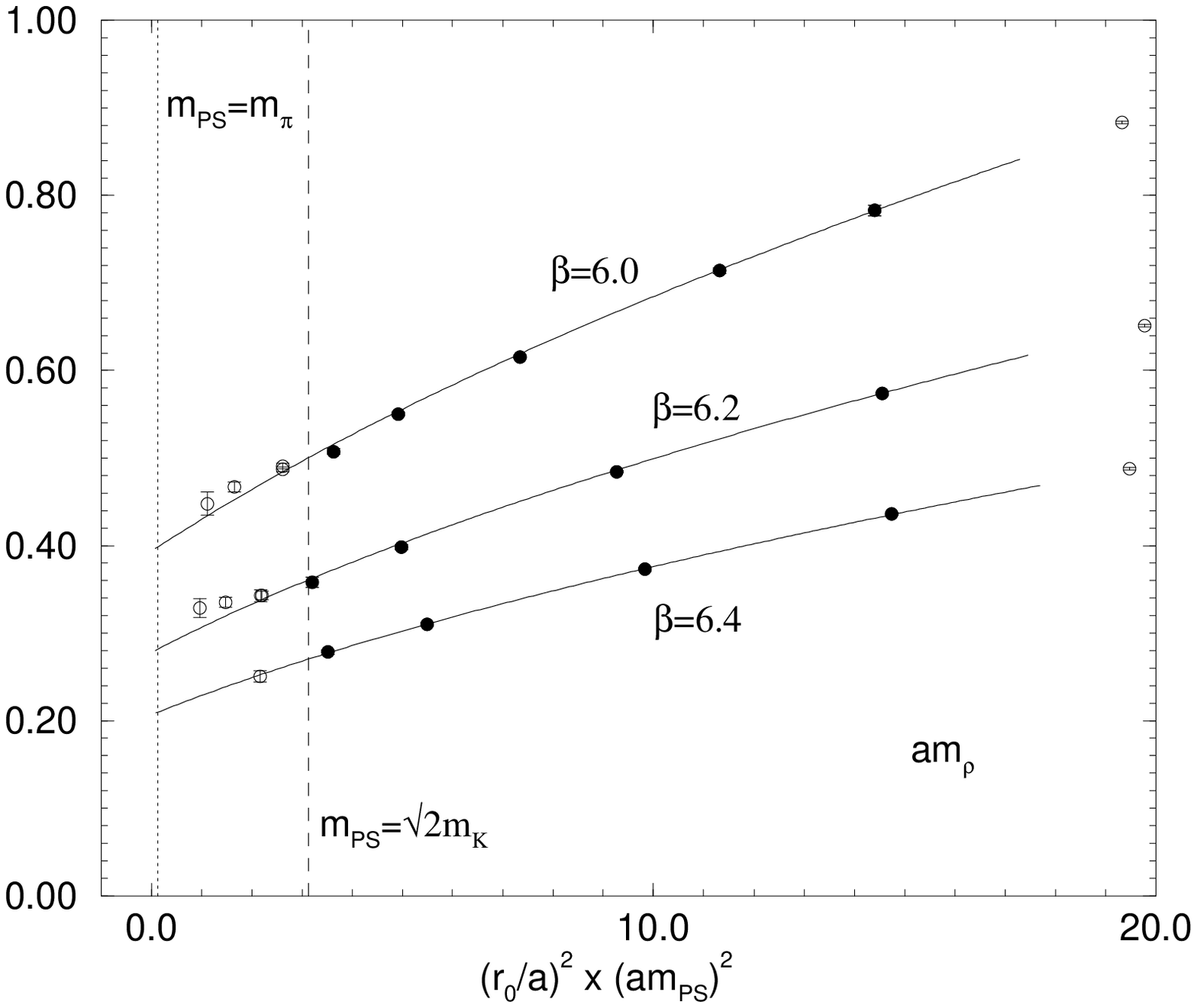}
   \caption{\it $am_\rho$ versus the pseudoscalar mass.
                The same notation
                as for Fig.~\ref{fig_mqwiompi2_mpi2_magic_lat99}.
                The fit function is given in
                eq.~(\ref{quenched_phenom_fit}).}
   \label{fig_rho_spectrum_dirk}
\end{figure}

\begin{figure}[htb]
   \hspace*{0.75in}
   \epsfxsize=8.50cm
   \epsfbox{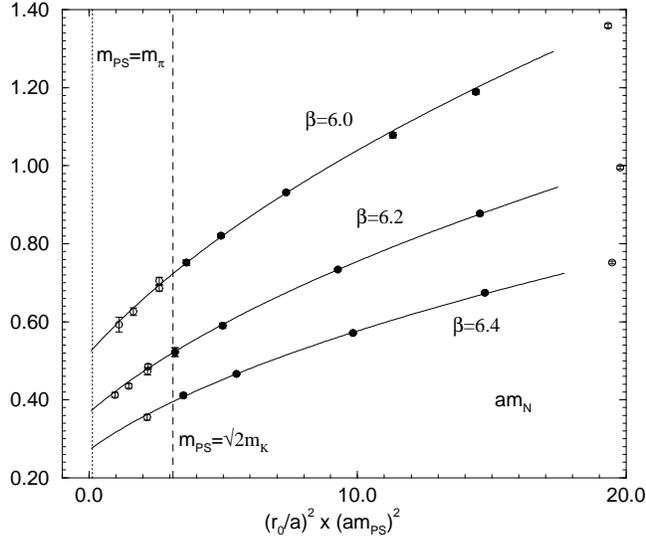}
   \caption{\it $am_N$ versus the pseudoscalar mass.
                The same notation
                as for Fig.~\ref{fig_rho_spectrum_dirk}.}
   \label{fig_nucl_spectrum_dirk}
\end{figure}

Finally we consider chiral extrapolations of the nucleon and rho
masses. In Figs.~\ref{fig_rho_spectrum_dirk}, \ref{fig_nucl_spectrum_dirk}
we show the results, together with a phenomenological fit
\begin{equation}
   (am_{\rho,N})^2 = b_0 + b_2 (a m_{ps})^2 + b_3 (a m_{ps})^3.
\label{quenched_phenom_fit}
\end{equation}
In distinction to eq.~(\ref{quenched_theory_fit}) this does
not have a quenched linear chiral term. (As there is curvature
in the results above the strange quark mass, we have considered
$(am_{\rho,N})^2$ rather than $am_{\rho,N}$ and included
a cubic term in eq.~(\ref{quenched_phenom_fit}).
This gave a better fit function for the data.)
While, for the nucleon this gives a good description of the data
over the whole quark mass range and it is thus difficult to say in
this case whether a linear term is necessary or not,
the $\rho$ data might be showing some deviations for small quark masses.

%----------------------------------------------------------------------------

\section{Conclusions}

Our main conclusion is that in quenched QCD there seems to be a
dangerous region for quark masses $m_q \lsim m_s$.
If we are interested in the strange quark mass or
particles such as $m_K$, $m_{K^*}$, $\ldots$, or decay constants such as
$f_K$, $f_{K^*}$, $\ldots$, this does not represent a problem.
For quantities involving only the $u$ and $d$ quarks, it
is probably best to adopt the pragmatic approach of making
fits for $m_q \gsim m_s$ and then to extrapolate this
to $m_q \sim m_{u/d}$, ie to chiral limit. (See, for example,
the results for the quark mass, \cite{gockeler99a}.)
Evidence for chiral quenched logarithms is mixed -- the best
signal seems to be for the pion and its associated decay
constant. Other channels seem to be less unambiguous.
Indeed, as detecting and measuring quenching effects
can be quite difficult
this would indicate that the quenched approximation often
seems to be working quite well.

The above results should be regarded as preliminary.
We hope to present full results shortly, \cite{gockeler00a},
including continuum extrapolations (considered in the talk,
but not described here).

%----------------------------------------------------------------------------

\section*{Acknowledgements}

The numerical calculations were performed on the
Quadrics {\it QH2} at DESY (Zeuthen) as well as
the Cray {\it T3E} at ZIB (Berlin) and the Cray {\it T3E} at
NIC (J\"ulich). We wish to thank all institutions for their support.

%----------------------------------------------------------------------------

%----------------------------------------------------------------------------

\end{document}